\documentclass{wsprocs}

\begin{document}

\title{Self-action effects in the theory of classical spinning charge
\footnote{\uppercase{W}ork partially
supported by grants 00-15-96566 and
\uppercase{LSS}-1578.2003.2 of the \uppercase{RFBR}.}}

\author{S.~L. Lebedev }
\address{Chuvash State Pedagogical University,\\
K.~Marx str. 38, \\
Cheboksary, 428000, Russia\\
E-mail:sll@chgpu.cap.ru  lsl@chuvsu.ru}


\maketitle

\abstracts{The back-reaction effects for the spinning charge moving through
the constant homogeneous electromagnetic field are studied in the context of
the mass-shift (MS) method. For the g=2 magnetic moment case we find the
(complex) addition to the classical action. Its dependence on the integrals
of the unperturbed motion proves to be important in determination of the
orbital radiation effects and could assist in understanding the radiation
polarization (RP) phenomenon.}

\section{Introduction}
The topic of this note is self-interacting classical charge possessing magnetic
moment. The recently renewed interest in the pseudoclassical models of spinning
particles stems from the close relations between those models and string
theory. With rare exception, the problem of self-interaction for the
spinning charge have not been considered there. At that time this problem
could find an interesting application in the theory of RP phenomenon \cite{Ba,Ter}.


The effects of self-interaction are usually approached through the
Abraham-Lorentz-Dirac (ALD) equation describing the radiation effects for the
spinless charge. Being generalized on a non-zero spin case, this approach
leads to inappropriately complicated equations (see e.g. \cite{Row}). On
the other hand, to analyse the self-action effects one can use the complex
addition \footnote{With obvious exceptions we use the system of units with $c=1,\, \hbar =1, \alpha
= e^{2}/4\pi \hbar c$, and 4-vector notations $x_{\mu}=({\bf x},ix_{0})$.}
$$
\Delta W=\left.\frac12 \int\int\,J_{\mu}(x)\Delta_{c}(x,x';\mu_{ph})J_{\mu}(x')
\,dx\,dx'\right |_{0}^{F}
\eqno{(1)}
$$
to the classical action functional of the particle. Inspired by QFT, this
approach \cite{Rit78,Rit81} relies on the fact (see e.g. \cite{IZ}) that
$\exp(\frac{i}{\hbar}\Delta W)$ is an amplitude and
$\exp(-\frac{2}{\hbar}\Im\Delta W)$ is the corresponding probability of the
photon vacuum to preserve when the classical source $J_{\mu}$ is present.
The causality of the Green function (GF) $\Delta_c$ in (1) guarantees the
account of radiation ($\Im\Delta W>0$) emitted by the source $J_{\mu}$. The
dependence of the self-action $\Delta W$ on the integrals of the unperturbed
motion carries an important information. For example, after specific procedure
of renormalization \cite{lsl2}, one can obtain an exact solution of the ALD
equation for the non-relativistic cyclotron motion in the following form:
$$
v_x+iv_y=Ae^{-i\Omega t}, \, \Omega=\frac{eH}{m+\delta m}, \,
\frac{\delta m}{m}=-\frac12+\sqrt{\frac14 +\frac{i\omega_c}{b}},\,
\frac1b=\frac23 \frac{e^2}{4\pi m}.
\eqno{(2)}
$$
Here $\omega_{c}=eH/m$ is the cyclotron frequency without regard for radiation
and $v_{x},v_{y}$ are velocity's components orthogonal to the magnetic field
$H$. The negativeness of $\Im\Omega$ originates from the causality of the
GF $\Delta_c$, and the dependence $\Delta W$ on the (unperturbed by radiation)
$v^{2}_{\perp}$ was the starting point.

This article discusses the simplified version of the polarization effects
for the spinning particle with no anomalous magnetic moment moving in
the constant homogeneous magnetic or electric field. For such external fields
the self-action $\Delta W$ reduces to the MS according to \cite{Rit81}
$$
\Delta W = -\Delta m \,T \, ,
\eqno{(3)}
$$
where the proper time $T$ corresponds to the interval of the charge's stay
in the external field. For the eq. (3) were meaningful the formation time
of the $\Delta m$ should be much less than $T$.

The need in adequate quasiclassical interpretation of the RP was pointed
out in the book \cite{Ter}. The wanted explaination would be done i)for the
different polarizations of electrons and positrons, ii) for not complete
(i.e. $<$100 p.c.) polarization degree (QED gives 0.924) and iii) for the
numerical value of the polarization time \cite{Ba,Ter}
$$
T_{QED}= \frac{8\sqrt{3}}{15}\frac{a_B}{c}\gamma_{\perp}^{-2}
\left(\frac{H_c}{H}\right)^3 \, ,
\eqno{(4)}
$$
($a_{B}=4\pi\hbar^2 /me^2, \,H_c= m^2c^3/e\hbar \sim 4.4 \,10^{13} Gs,
\gamma_{\perp}^{-2}=1-v^{2}_{\perp}$). An elementary classical consideration
\cite{Ter} leads within a factor of order 1 to the same value for the
characteristic time $T_{QED}$. This shows that quantum nature of RP might
be associated with the relationship $T_{QED} \sim 1/\mu_B$ only ($\mu_B$ being
the Bohr magneton) and needs quantum description neither for the orbital
motion \cite{Ba} nor for the spin precession.
\section{The general formulae}
The source $J_\alpha$ in (1) consists of orbit part
$$
j_{\alpha}(x)=e\,\int d\tau \dot x_{\alpha}(\tau)\delta^{(4)}(x-x(\tau))
\eqno{(5)}
$$
and spin contribution $\partial_{\beta}M_{\alpha\beta}(x)$, where
$$
M_{\alpha\beta}(x)=\int\, d\tau \mu_{\alpha\beta}(\tau)\delta^{(4)}(x-x(\tau))
\eqno{(6)}
$$
is the polarization density. The dependence of
$$
\mu_{\alpha\beta}=i\mu\varepsilon_{\alpha\beta\gamma\delta}\dot
x_{\gamma}S_{\delta} \eqno{(7)}
$$
on $\tau$ is determined from the Lorentz and Bargmann-Michel-Telegdi (BMT)
equations (see e.g. \cite{IZ}):
$$
\ddot x_{\alpha}=\frac{e}{m}F_{\alpha\beta}\dot x_{\beta},\,\,\,
\frac{\hbar}{2}\dot S_{\alpha}=\mu F_{\alpha\beta}S_{\beta}+(\frac g2-1)\mu_{B}
\dot x_{\alpha}(\dot x\cdot F\cdot S) .
\eqno{(8)}
$$
Here $\mu =\frac g2\mu_{B},\,\,S_{\mu}S_{\mu}={\bf \zeta}^2 = 1$, the overdots
denote the derivatives w.r.t. proper time $\tau$, and, in what follows, we
put $g=2$.

After substitution of the source $J_{\mu}=j_{\mu}+\partial_{\nu}M_{\mu\nu}$
in the r.h.s. of the eq.(1) and integration by parts, we find that
$$
\Delta W=\Delta W_{or}+\Delta W_{so}+\Delta W_{ss}\, .
\eqno{(9)}
$$
The orbit part $\Delta W_{or}=-\Delta m_{or}\cdot T$ for the electric or
magnetic external fields was considered in \cite{Rit81} and will not be
discussed below. The "spin-orbit" and "spin-spin" terms are:
$$
\Delta W_{so}=\left.-e\int d\tau \int d\tau'\dot x_{\beta}(\tau)
\,\mu_{\beta\alpha}(\tau ')\partial_{\alpha}^{'}\Delta_c(x,x';\mu_{ph})\right
|^F_0 \,,
\eqno{(10)}
$$

$$ \Delta W_{ss}=\left.\frac12 \int d\tau \int d\tau '\mu_{\alpha\beta}\,
\mu_{\alpha\gamma}^{'}\partial_{\beta}\partial_{\gamma}^{'}
\Delta_c(x,x';\mu_{ph})\right |^F_0 \,.
\eqno{(11)}
$$
The spin-orbit and spin-spin terms in eq.(9) form the small corrections to
orbital one. For example, the magnetic MS ratio $\Delta m_{so}/\Delta m_{or}
\simeq (H/H_c)\gamma_{\perp}$ \cite{lsl2}, so that only in the far quantum
region those terms could be of the same order. Below we shall focus our
attention on $\Delta m_{so}$ only (see (3)), regarding the latter as a major
contribution w.r.t. $\Delta m_{ss}$. Since infrared regulator $\mu_{ph}$
as well as the subtraction $|_{0}^{F}$ could be omitted here, we arrive at
$$
\Delta W_{so}=-\frac{\mu e}{2\pi^2}\int d\tau \int d\tau'
\frac{\varepsilon_{\alpha\beta\gamma\delta}(x-x')_{\alpha}\dot x_{\beta}(\tau)
\dot x_{\gamma}(\tau')S_{\delta}(\tau')}{[(x-x')^2]^2} \, ,
\eqno{(12)}
$$
where the use was made of
$$
\Delta_c(x,x';0)=i(2\pi)^{-2}/[(x-x')^2+i0] \, .
\eqno{(13)}
$$
\section{The spinning charge in magnetic field}

Choosing the direction of ${\bf H}$ along the z-axis, ${\bf H}=(0,0,H)$,
we have the following integrals of the motion: z-component of the
four-velocity $u_{3}=v_3\gamma$ and corresponding spin component $S_3$; the
energy $mu_0$, the scalar product ${\bf S_{\perp}\cdot v_{\perp}}$ of
vectors orthogonal to ${\bf H}$ and $v^2_{\perp}$. With
$u^2_{\perp}=u^2_0-u^2_3-1\equiv u^2_{\|}-1=v^2_{\perp}\gamma^2,
\gamma=(1-{\bf v}^2)^{-1/2},\,\,{\bf v}^2=v^2_3+v^2_{\perp}$ and after some
algebra the expressions (3) and (12) give rise to:
$$
\Delta m_{so}=-i\frac{\mu e}{2\pi^2}(S_3-v_3 S_0)\,\omega_c^2
f_m(v_{\perp},v_3)\, ,
\eqno{(14)}
$$
where formfactor
$$
f_m(v_{\perp},v_3)=2v^2_{\perp}\gamma^3\int^{\infty}_{0}\,
\frac{4\sin^2{(x/2)}-x\sin{x}}{(4u^2_{\perp}\sin^2{(x/2)}-u^2_{\|}x^2)^2}\,dx
\eqno{(15)}
$$
takes the retardation effects into account (variable
$x=\omega_c(\tau-\tau')$). Note, that everywhere we put $e=-|e|$ and that,
according to "Frenkel condition",
$$
S_0={\bf S\cdot v} \, .
\eqno{(16)}
$$
In calculating the determinant present as a nominator of the integrand in eq.
(12), one finds \footnote{$u_{[0}S_{3]}$ means antisymmetrized combination
$u_0S_3-u_3S_0$.}
$$
\varepsilon_{\alpha\beta\gamma\delta}(x-x')_{\alpha}\dot x_{\beta}(\tau)
\dot
x_{\gamma}(\tau')S_{\delta}(\tau')=-iu^2_{\perp}u_{[0}S_{3]}\,\omega_c^{-1}
(4\sin^2{(x/2)}-x\sin{x})\, , \eqno{(17)} $$
so that $\Delta m_{so}$ would have an opposite sign for positrons because of
the factor $\omega_c^{-1}$.

Considering experimental situation \cite{Ter} we put $v_3=0$ and find that
spin contribution into the amplitude of the vacuum preservation reduces to the
factor $\exp{(\Im \Delta m_{so}\cdot T)}$ where, in accordance with (14),
$\Im \Delta m_{so}$ is positive when $S_3 <0$ and negative otherwise. Hence,
the probability of radiation from `spin-down` electron as compared with
`spin-up` one is by the factor
$$
\frac{\exp{(2\Im \Delta m_{or}T+2\Im \Delta m_{so}(\uparrow)T)}}
{\exp{(2\Im \Delta m_{or}T+2\Im \Delta m_{so}(\downarrow)T)}}=
\exp{(-4T\frac{\mu e}{2\pi^2}\omega_c^2f_m)}
\eqno{(18)}
$$
suppressed. To obtain the same factor in the positron case one should inverse
the directions of arrows in the l.h.s. of eq.(18). The characteristic
laboratory time $T_{char}$ which is deduced from (18),
$$
T_{char}\simeq \left(\frac{\mu e}{\pi^2}\,\omega_c^2f_m
\right)^{-1}\gamma_{\perp}=2\sqrt{3}\,\frac{a_B}{c}\gamma_{\perp}^{-1}
(H_c/H)^2 \, ,
\eqno{(19)}
$$
differs\footnote{The last expression accounts for the relativistic
asymptotics of $f_{m}(v_{\perp},0)$ \cite{lsl2}.} from $T_{QED}$ in (4)
(numerical evaluation for $H\sim 10^4 Gs$ and electron energy about $1 GeV$
shows $T_{QED}\sim 5\cdot 10^5 T_{char}$). It should not be considered as a
surprise because time $T_{char}$ accumulate all of the possibilities for
electron to leave the initial state, so that it {\it should} be less than
$T_{QED}$. Note that formation time of $\Delta m_{so}$ (extracted from (15)
at $v_3=0$) is $x/\omega_c =\Delta \tau\sim \omega_c^{-1}\gamma_{\perp}^{-1}
$, and $T_{char}/\gamma_{\perp}\Delta\tau \simeq \gamma_{\perp}^{-1}
(H_{c}/\alpha H) \gg 1$ (see the text following the eq.(3)).
\section{The spinning charge in electric field}
The list of the integrals of the motion in electric field ${\bf E}=(0,0,E)$
is following: $u_1,\, u_2,\, S_1,\,S_2$ being the $(x,y)$-components of the
4-velocity $\dot x_{\mu}$ and the spin $S_{\mu}$ respectively. The substitution
of the solutions of eqs.(8) ($g=2,\, e=-|e|$) into r.h.s. of (12) leads to
$$
\Delta m_{so} = -i\frac{\mu e}{2\pi^2}\, w^2\,\bar v_{[1}S_{2]}f_e(u^2_{\perp})\, ,
\eqno{(20)}
$$
$$
f_e(u^2_{\perp})=2\gamma_0^3\int^{\infty}_{0}\,
\frac{x\sinh{x}-4\sinh^2{(x/2)}}{[u^2_{\perp}x^2-4u^2_{\|}\sinh^2{(x/2)}]^2}\,dx ,
\eqno{(21)}
$$
with $w=eE/m,\, \gamma_0^2=1+u^2_{\perp}=u^2_{\|},\, u^2_{\perp}=u^2_1+u^2_2 $.
$\bar v_{1,2}\equiv u_{1,2}/\gamma_0 $ are the $(x,y)$ - components of velocity
${\bf v}$ taken at the moment when the component $v_3=0$.

The wanted asymmetry between polarizations is associated with the factor
${\bf(\bar v\times S)\cdot E}$. One can see that $\Im \Delta m_{so}>0 $ when
${\bf(\bar v\times S)}_3<0$, i.e. when the system ${\bf{S,\bar v,E}}$ forms the
right-handed triple. This orientation corresponds to the suppressed radiation
probability by the factor $\exp{(-2\Im \Delta m_{so}T)}$ (cf. with (18)). For
positrons the primary direction is, of course, opposite. Notice that orbital
part of the MS in electric field involves the infrared
singularity signalling about the large formation time of the MS \cite{Rit81}.
This does not affect the relative quantities like l.h.s of eq. (18), but makes
it necessary to weaken an `orbital background` of the spin radiation \cite{Ter}.
That could in principle be done by the employment of boundaries
\cite{lsl94}.

\section{Conclusion}
The primary purpose of this note was to demonstrate a new possibility to
account for radiation effects in the dynamics of the spinning charge. The
present consideration is certainly not complete. One could ask about the
role of $(g-2)$- term in BMT equation, see (8). It is known \cite{Ba} that
in the relativistic limit this term might be of first importance. The next
point is $\Delta W_{ss}$ in (11) which we are going to discuss in a subsequent
publication. Its dependence on $S_3$ would be expected to clarify the
non-complete polarization degree\footnote{Qne can guess that the retardation
interaction term $\propto \mu_{\alpha\beta}\,\mu_{\alpha\gamma}^{'}$ in (11)
corresponds to a recoil effects of spin radiation.}. Of some practical interest
could be the exact dependences of the formfactors $f_m$ and $f_e$ on dynamical
invariants as well as not discussed yet the possibility to observe RP in
electric field.
\section{Aknowledgements}
The author is grateful to RFBR for partial financial support
(grants 00-15-96566 and LSS-1578.2003.2)


\end{document}